# Copyright Laundering Through the AI Ouroboros: Adapting the 'Fruit of the Poisonous Tree' Doctrine to Recursive AI Training

Anirban Mukherjee

Hannah Hanwen Chang

Anirban Mukherjee (anirban@avyayamholdings.com) is Principal at Avyayam Holdings. Hannah H. Chang (hannahchang@smu.edu.sg; corresponding author) is Associate Professor of Marketing at the Lee Kong Chian School of Business, Singapore Management University. This research was supported by the Ministry of Education (MOE), Singapore, under its Academic Research Fund (AcRF) Tier 2 Grant, No. MOE-T2EP40221-0008.

**Abstract**

Copyright enforcement rests on an evidentiary bargain: a plaintiff must show both the defendant's access to the work and substantial similarity in the challenged output. That bargain comes under strain when AI systems are trained through multi-generational pipelines with recursive synthetic data. As each successive model is tuned on the outputs of its predecessors, any copyrighted material absorbed by an early model is further diffused into deeper statistical abstractions. The result is an evidentiary blind spot where overlaps that emerge look coincidental, while the chain of provenance is too attenuated to trace.

These conditions are ripe for "copyright laundering"—the use of multi-generational synthetic pipelines, an "AI Ouroboros," to render traditional proof of infringement impracticable. This Article adapts the "fruit of the poisonous tree" (FOPT) principle to propose a novel AI-FOPT standard: if a foundational AI model's training is adjudged infringing (either for unlawful sourcing or for non-transformative ingestion that fails fair-use), then subsequent AI models principally derived from the foundational model's outputs or distilled weights carry a rebuttable presumption of taint. The burden then shifts to the downstream developers—those who control the evidence of provenance—to restore the evidentiary bargain by affirmatively demonstrating a verifiably independent and lawfully sourced lineage or a curative rebuild, without displacing fair-use analysis at the initial ingestion stage. Absent such proof, commercial deployment of tainted models and their outputs is actionable.

Drawing on legal precedents, this Article develops the standard by specifying its trigger, presumption, and concrete rebuttal paths (such as independent lineage or verifiable unlearning); addresses counterarguments concerning chilling innovation and fair use; and demonstrates why this lineage-focused approach is both administrable and essential to preserve copyright's efficacy and incentive structure when machines train machines.

---

# Introduction

Copyright liability has always been evidence-driven. A plaintiff prevails only by showing that the defendant (i) had *access* to a protected work and (ii) reproduced a *substantial similarity* of its protectable expression.[1] That doctrinal bargain allows courts to calibrate owners' incentives against follow-on creativity: a bright-line "no sampling" rule when even a two-second riff is recognizable,[2] a forgiving standard when wholesale copying serves a radically different purpose and unmistakably transforms the original.[3] Either way, the inquiry has been possible because there is something concrete—text, image, track—against which the new work can be compared.

The use of synthetic[4] data in training generative artificial intelligence (AI) now threatens to break that evidentiary bargain. Imagine an initial foundation model ($AI_1$), trained on a gargantuan corpus that indisputably contains copyrighted works. Its outputs are *scrubbed*—that is, filtered or processed to remove overtly infringing content—and re-packaged as synthetic datasets used to train a successor model;[5] that successor, in turn, could inform its own successor, and so on: an *AI Ouroboros* that buries the original human source material ever deeper beneath opaque layers of statistical abstraction.[6]

---

[1] Arnstein v. Porter, 154 F.2d 464, 468-69 (2d Cir. 1946); Nichols v. Universal Pictures Corp., 45 F.2d 119, 121 (2d Cir. 1930).

[2] Bridgeport Music, Inc. v. Dimension Films, 410 F.3d 792, 801–02 (6th Cir. 2005).

[3] Campbell v. Acuff-Rose Music, Inc., 510 U.S. 569, 579 (1994).

[4] Synthetic data refers to information that is algorithmically generated to emulate the statistical properties of real-world datasets, rather than being directly observed or authored by humans. *See, e.g.*, James Jordon et al., *Synthetic Data: What, Why And How?* 5–7, Roy. Soc'y (2022)(providing a foundational overview of synthetic data); Yingzhou Lu et al., *Machine Learning for Synthetic Data Generation: A Review*, arXiv:2302.04062, at 1-2 (v10 Apr. 4, 2025), https://arxiv.org/abs/2302.04062 [https://perma.cc/7CTG-LFTH] (surveying generation techniques and definitions, and noting its use to "augment or replace real data in various AI tasks"). It is distinct from *human-generated* data (e.g., original literary works or artistic creations) and *human-annotated* data (where humans provide labels or categorizations for existing content).

[5] The use of AI-generated outputs as training material (synthetic data) for successor models is increasingly prevalent and is gaining recognition in regulatory discussions concerning AI transparency and data provenance. *See, e.g.*, U.S. Copyright Office, *Copyright And Artificial Intelligence: Part 3—Generative Ai Training* § II.D (May 2025) (Pre-Publication Version) (discussing various data sources for AI training, including synthetic data generated by other AI, and the complexities this introduces for copyright); Regulation (EU) 2024/1689 of the European Parliament and of the Council of 13 June 2024 laying down harmonised rules on artificial intelligence (Artificial Intelligence Act), art. 53(1)(d), 2024 O.J. (L, 2024/1689) 1 (requiring providers of general-purpose AI models to "draw up and make publicly available a sufficiently detailed summary about the content used for training" the model); *see id.* Recitals (107)–(110).

By the time a late-generation model produces user-facing content, any overlap with the original corpus appears merely coincidental—the "infinite-monkey" problem amplified,[7] its lineage buried deep within billions of parameters[8] and, in cutting-edge systems, dispersed across mixture-of-experts sub-models with distinct training histories.[9]

---

[6] While naively training AI models exclusively on their own prior outputs can lead to performance degradation or "model collapse," *see, e.g.*, Ilia Shumailov et al., *AI Models Collapse When Trained on Recursively Generated Data*, 631 NATURE 755–759 (2024) (Author Correction Mar. 21, 2025), this does not preclude the "AI Ouroboros" scenario. The concern remains that an initial model ($AI_1$), trained on human-authored (and potentially copyrighted) material, can have its outputs form a substantial basis for training subsequent models ($AI_2$, etc.). Indeed, emerging techniques demonstrate that successor models (like $AI_2$) can be effectively trained or refined using synthetic data generated by a precursor model (like $AI_1$), coupled with feedback mechanisms (human or AI-driven) to guide development and potentially avoid such collapse. *See, e.g.*, Yuntao Bai et al., *Constitutional AI: Harmlessness from AI Feedback*, ARXIV:2212.08073 (Dec. 15, 2022), https://arxiv.org/abs/2212.08073 [https://perma.cc/5FYV-HB4X] (utilizing AI-generated feedback for model alignment); Harrison Lee et al., *RLAIF vs. RLHF: Scaling Reinforcement Learning from Human Feedback with AI Feedback*, ARXIV:2309.00267 (Sept. 1, 2023), https://arxiv.org/abs/2309.00267 [https://perma.cc/XJ47-5HTW] (demonstrating AI-generated feedback in reinforcement learning); Lewis Tunstall et al., *Zephyr: Direct Distillation of LM Alignment*, ARXIV:2310.16944 (Oct. 25, 2023), https://arxiv.org/abs/2310.16944 [https://perma.cc/F6S6-XK3Y] (distilling alignment preferences from a more capable AI model). If such methods are employed—where $AI_1$'s outputs (potentially filtered or refined by such feedback) train $AI_2$—or if future systems achieve greater autonomy through pure reinforcement learning (as demonstrated in closed domains, *see, e.g.*, David Silver et al., *A General Reinforcement Learning Algorithm that Masters Chess, Shogi, and Go Through Self-Play*, 362 SCIENCE 1140–1144 (2018))—then the influence of $AI_1$'s potentially infringing training may persist and be "laundered" through a lineage of successive, synthetically-derived AI model generations.

[7] The "infinite-monkey" problem refers to the theoretical notion that given infinite time, a monkey randomly typing on a typewriter would eventually produce any given text, such as a Shakespearean sonnet. Modern generative AI reasoning models amplify this problem significantly. These models are explicitly trained to recognize and select outputs based on certain quality metrics—such as coherence, style, or literary merit. Thus, even if the underlying generative process is initially random or only weakly guided, the reasoning model will systematically favor outputs that closely resemble high-quality human-authored texts (such as a Shakespearean sonnet). This selective pressure dramatically increases the likelihood that a model will produce outputs that appear coincidentally similar to copyrighted works, making it practically impossible to distinguish genuine coincidence from infringement. *See, e.g.*, Long Ouyang et al., *Training Language Models to Follow Instructions with Human Feedback*, NEURIPS 2022, at 1–2, https://arxiv.org/abs/2203.02155 [https://perma.cc/CY7S-AQDT] (showing that preference-optimized training concentrates output probability mass on human-preferred text); Daniel M. Ziegler et al., *Fine-Tuning Language Models from Human*

These complexities render direct forensic detection practically impossible. Courts relying on side-by-side comparison are left with no discernible artifact to analyze and no principled way to distinguish a coincidental echo from an original, yet "laundered," copy. This creates an evidentiary blind spot, fostering conditions ripe for what this Article terms "copyright laundering"—the use of multi-generational, synthetic-data training pipelines to systematically erase provenance by diffusing protectable expression across successor training corpora until traditional access-and-similarity proofs become forensically impracticable.[10]

Early litigation highlights the evidentiary challenge. In *Thomson Reuters Enter. Ctr. GmbH v. ROSS Intelligence Inc.*, the court could still trace Westlaw's headnotes to an AI system trained to answer legal queries; the linkage was concrete and the use was not transformative.[11] Yet this methodology falters when an author seeks to sue an AI model *n* generations removed from the model that directly ingested her novel.[12] Traditional tests—access plus similarity—

---

*Preferences*, ARXIV:1909.08593 (last revised Jan. 8, 2020), at 1–2, https://arxiv.org/abs/1909.08593 [https://perma.cc/XPA5-A52Q].

[8] This statistical abstraction is a core challenge. Generative AI models typically do not store or reproduce training data verbatim but rather learn statistical patterns and relationships, which are then encoded as parameters. *See* Pamela Samuelson, *Generative AI Meets Copyright*, 381 SCIENCE 158–61 (2023) (noting that Stable Diffusion "contains an extremely large number of parameters that mathematically represent concepts embodied in the training data, but the images as such are not embodied in its model," and that training "uses these tokens to discern statistical correlations—often at staggeringly large scales").

[9] *See generally* Noam Shazeer et al., *Outrageously Large Neural Networks: The Sparsely-Gated Mixture-of-Experts Layer*, ARXIV:1701.06538 (Jan. 23, 2017), https://arxiv.org/abs/1701.06538 [https://perma.cc/NY3W-N599].

[10] The challenges posed by generative AI to traditional infringement analysis are significant. *See, e.g.*, Mark A. Lemley, *How Generative AI Turns Copyright Law Upside Down*, 25 COLUM. SCI. & TECH. L. REV. 190 (2024) (explaining how AI's "many-inputs-to-one-output" model breaks the traditional evidentiary link between a specific copyrighted work and a specific infringing output).

[11] Thomson Reuters Enter. Ctr. GmbH v. Ross Intel. Inc., 765 F. Supp. 3d 382, 397–99 (D. Del. 2025) (granting partial summary judgment; rejecting fair use defense); *see also, e.g.,* Complaint ¶¶ 51–64, Getty Images (US), Inc. v. Stability AI, Inc., No. 1:23-cv-00135 (D. Del. filed Feb. 3, 2023); Complaint ¶¶ 59–60, N.Y. Times Co. v. Microsoft Corp., No. 1:23-cv-11195 (S.D.N.Y. filed Dec. 27, 2023) (alleging that defendants concealed training data composition from rightsholders).

[12] This is increasingly the case whereby publicly available AI are late-generation models that have evolved over time. For example, OpenAI's GPT models (GPT-3, GPT-3.5, GPT-4, GPT-4.5, GPT-4o, etc.) and Google's Gemini models (Gemini 1, Gemini 1.5, Gemini 2, Gemini 2.5,

impose a prohibitive burden on the author. She must prove not only what was in the original (potentially opaque) training dataset,[13] but also trace how that material persisted through multiple rounds of synthetic regeneration.

This Article argues that the only doctrinally workable response[14] to this evidentiary crisis is to adapt the "fruit of the poisonous tree" (FOPT) doctrine. No output-focused standard can solve an input-erasure tactic; no expansion of copyright's subject matter can survive the idea/expression dichotomy; and no transparency-only regime reallocates the prohibitive burden of proof. What remains is an evidentiary rule calibrated to *lineage* rather than *likeness*.

The FOPT doctrine is uniquely suited to this task precisely because it reallocates the burden of proof to the party with unique and low-cost access to the dispositive evidence of a model's provenance.[15] Specifically, if an initial model's ($AI_1$) training is adjudged infringing—

---

etc.) are iteratively trained, fine-tuned, and updated, progressively obscuring the original training data lineage. *See, e.g.*, OpenAI, GPT-4 Technical Report 2 (2023) (describing GPT-4 as "pre-trained to predict the next token in a document, using both publicly available data (such as internet data) and data licensed from third-party providers," without disclosing dataset construction details); *see also* Ilia Shumailov et al., *AI Models Collapse When Trained on Recursively Generated Data*, 631 Nature 755, 755–59 (2024).

[13] Doctrinally, the "access" prong asks only for a reasonable possibility that defendants could have obtained the work. *See, e.g.*, Towler v. Sayles, 76 F.3d 579, 582 (4th Cir. 1996); Bouchat v. Baltimore Ravens, Inc., 241 F.3d 350, 354-55 (4th Cir. 2001) (applying the "reasonable possibility of access" standard). In large-scale AI scraping, however, defendants can contend that any overlap is coincidence and insist on proof of actual ingestion to defeat summary judgment. Thus, the practical burden on plaintiffs often balloons from showing *feasible* access to demonstrating *actual* inclusion and persistence through successive model generations. *See* Cong. Rsch. Serv., LSB10922, *Generative Artificial Intelligence and Copyright Law* 5 (July 18, 2025) ("For AI outputs, access might be shown by evidence that the AI program was trained using the underlying work."); *see also* Andersen v. Stability AI Ltd., No. 23-cv-00201-WHO, 2023 WL 7132064, at *6–7 (N.D. Cal. Oct. 30, 2023) (allowing direct infringement claim to proceed where plaintiff pointed to haveibeentrained.com search results to identify works in training data).

[14] By "doctrinally workable," this Article means a judge-made approach that (i) fits within a court's existing equitable and evidentiary powers without creating new exclusive rights; (ii) restores a realistic evidentiary pathway for plaintiffs notwithstanding the opacity of synthetic-data pipelines; and (iii) remains cabined by familiar limiting principles, such as rebuttable presumptions and proportional remedies.

[15] While scholars have cautioned against a rigid FOPT doctrine in intellectual property generally, arguing it could chill legitimate follow-on creation, the unique evidentiary pathologies of multi-generational AI call for a carefully calibrated version of that principle specifically tailored to this new context. *See, e.g.,* Mark A. Lemley, *The Fruit of the Poisonous Tree in IP Law*, 103 IOWA L. REV. 245, 263–65 (2017) (cautioning that a strict FOPT could penalize valuable non-infringing downstream products). We address these concerns more fully in Part II and Part IV, arguing that our AI-specific proposal mitigates many of the risks identified in broader IP applications.

because the corpus was copied without authorization or fails fair-use scrutiny—[16]every dataset and model principally derived[17] from its outputs ($AI_2$, $AI_3$, and so on) carry a rebuttable presumption of taint. The burden then shifts to the developer of $AI_1$'s successor models ($AI_2$, $AI_3$, …) to demonstrate a *verifiably independent* and lawfully sourced lineage. Absent that, commercial deployment of the tainted model (and its outputs) is actionable even when no side-by-side similarity can be marshaled.

The Article proceeds as follows. Part I reviews the doctrinal foundations of copyright's access-and-similarity paradigm. Part II explains why multi-generational, synthetic-data pipelines render that paradigm unworkable and demonstrates why other doctrinal tools are insufficient to address the problem. Part III develops the proposed AI-FOPT standard, drawing on precedents that already withhold legal benefit from downstream exploitation of an initial infringement. Part IV addresses anticipated objections—such as chilling innovation and conflicts with fair use—and demonstrates why a lineage-based rule is both administrable and essential to preserving copyright's efficacy and incentive structure in this age of machine-trained machines. Part V concludes.

---

[16] This Article brackets the merits of the *first* model's ingestion, the lawfulness of which rests on a two-part inquiry under existing doctrine—*sourcing* (were copies lawfully obtained) and *use* (is the purpose transformative and non-substitutive under § 107). A model is not a "poisonous tree" if its training on *lawfully obtained* copies is adjudged a *transformative* fair use for a distinct, non-substitutive purpose and the intermediate copies are neither distributed nor used to supplant the market for the originals. *See* Authors Guild v. Google, Inc., 804 F.3d 202, 214–19 (2d Cir. 2015); *see also* Bartz v. Anthropic PBC, 787 F. Supp. 3d 1007, 1023-25, 1031-32 (N.D. Cal. 2025) (Order on Fair Use). By contrast, a model becomes a "poisonous tree" if it fails on either sourcing or use. It fails the *use* prong if its ingestion, even of lawfully accessed materials, is for a non-transformative purpose that creates a market substitute for the original. *See* Thomson Reuters Enter. Ctr. GmbH v. Ross Intel. Inc., 765 F. Supp. 3d 382, 397-98 (D. Del. 2025) (Order Granting Partial Summary Judgment). It fails the *sourcing* prong if its training relies on unlawfully obtained materials—such as pirated corpora from Books3 or LibGen—retained "forever" in a central, general-purpose library; fair use does not excuse the initial piracy, and later purchase does not cure it. *See* Bartz v. Anthropic PBC, 787 F. Supp. 3d 1007, 1032-33 (N.D. Cal. 2025) (Order on Fair Use)(.

[17] "Principally derived," as used throughout this Article, refers to a downstream model or dataset's material dependence on a predecessor. This is a fact-intensive inquiry that can be assessed by quantitative measures (e.g., a substantial share of training data for $AI_2$ originates from $AI_1$'s outputs) and qualitative importance (e.g., $AI_2$ was initialized from $AI_1$'s distilled weights or fine-tuned on synthetic data that imparts a core capability). Part III develops this standard further.

# I. COPYRIGHT'S EVIDENCE MACHINE: THE PRESUMPTION OF VISIBILITY

This Part reviews the doctrinal foundations of copyright infringement, showing how its core requirements—*access* and *substantial similarity*—traditionally form an "evidence machine" premised on a single assumption: the *presumption of visibility*, that enables courts to adjudicate infringement through direct, tangible comparison of the original and the allegedly infringing work.

The initial threshold for any infringement claim is *access*: a plaintiff must show a reasonable possibility that the defendant had an opportunity to view or copy the original work.[18] While traditional cases grappled with evidence of sheet-music sales, broadcast reach, or manuscripts delivered to publishers,[19] modern courts often find constructive access where a work is widely available online.[20] Regardless of the context, access has functioned as a crucial gateway; without it, even uncanny similarities between works are typically attributed to coincidence, and the claim fails.[21] Once access is established—or, in rare cases, inferred from "striking similarity"—the inquiry turns to whether the defendant improperly appropriated protectable expression.[22]

The determination of *substantial similarity* has evolved through several strands. One prominent approach is the *ordinary observer* test, which asks whether a lay observer would regard the aesthetic appeal of the two works as the same, considering their "total concept and feel."[23] For more functional works, particularly software, courts developed the *Abstraction-Filtration-Comparison (AFC)* test, as articulated in *Computer Associates International, Inc. v. Altai, Inc.*[24] This methodical approach dissects works to filter out unprotectable ideas, processes, and *scènes à faire* before comparing the remaining protectable expression. Originally articulated for software, this framework has since been adopted by other circuits,[25] and its filtration logic has been applied in other contexts.[26]

---

[18] Arnstein v. Porter, 154 F.2d 464, 468-69 (2d Cir. 1946) ("[T]he evidence may consist of . . . circumstantial evidence —usually evidence of access—from which the trier of the facts may reasonably infer copying.").

[19] *See* Three Boys Music Corp. v. Bolton, 212 F.3d 477, 482 (9th Cir. 2000) (citing 4 Melville B. Nimmer & David Nimmer, *Nimmer on Copyright,* § 13.02[A], at 13-20-13-2 (1999); 2 Paul Goldstein, *Copyright: Principles, Law, and Practice* § 8.3.1.1., at 90–91 (1989)).

[20] *See, e.g.*, Art Attacks Ink, LLC v. MGA Entm't, Inc., 581 F.3d 1138, 1143–45 (9th Cir. 2009) (stating the widespread dissemination rule but finding access not established on the facts).

[21] Sheldon v. Metro-Goldwyn Pictures Corp., 81 F.2d 49, 54 (2d Cir. 1936).

[22] *See* Skidmore v. Led Zeppelin, 952 F.3d 1051, 1064, 1066–69 (9th Cir. 2020) (en banc) (rejecting the "inverse ratio" rule linking the quantum of similarity to the showing of access).

[23] Peter Pan Fabrics, Inc. v. Martin Weiner Corp., 274 F.2d 487, 489 (2d Cir. 1960) (articulating the influential "ordinary observer" test for aesthetic appeal); *see also* Roth Greeting Cards v. United Card Co., 429 F.2d 1106, 1110 (9th Cir. 1970) (coining "total concept and feel").

[24] Comput. Assocs. Int'l, Inc. v. Altai, Inc., 982 F.2d 693, 706–11 (2d Cir. 1992).

Copyright also protects against fragmented but qualitatively significant takings. Even a brief excerpt can infringe if it captures the "heart" of the original,[27] a principle that underpins the Second Circuit's protection of the whole dramatic meaning in *Sheldon v. MGM*[28] and, more controversially, the Sixth Circuit's more rigid, bright-line rule in *Bridgeport Music, Inc. v. Dimension Films*, which declared "no de minimis" for recognizable sound recording samples.[29]

Beyond direct reproduction of an entire work, copyright grants owners the exclusive right to prepare *derivative works*—transformations or adaptations of their original creations.[30] Here too, the analysis hinges on a side-by-side comparison: courts determine whether the allegedly derivative work incorporates a substantial quantum of the original's protectable expression, effectively "recasting, transforming, or adapting" that expression. For example, in *Castle Rock Entertainment, Inc. v. Carol Publishing Group, Inc.*, the Second Circuit found a "Seinfeld" trivia book to be an infringing derivative work because it systematically appropriated numerous protected elements, such as specific dialogue, characters, and plot details, from the original television series.[31] Similarly, under EU and UK law, the rights of reproduction and adaptation are infringed if a new work incorporates a "substantial part" of an original work, where substantiality is judged qualitatively on whether the part taken reflects the author's own intellectual creation.[32]

---

[25] *See, e.g.*, Engineering Dynamics, Inc. v. Structural Software, Inc., 26 F.3d 1335, 1342–44 (5th Cir. 1994), supplemented on reh'g, 46 F.3d 408, 409 (5th Cir. 1995) (endorsing the Abstraction-Filtration-Comparison approach); *see* also Kepner-Tregoe, Inc. v. Leadership Software, Inc., 12 F.3d 527, 536 & 536 n.20 (5th Cir. 1994) (recognizing protection for non-literal elements and approving filtration).

[26] Boisson v. Banian, Ltd., 273 F.3d 262, 272 (2d Cir. 2001) (using a "more discerning ordinary observer" that filters out unprotectable elements before comparing protectable expression in quilt designs).

[27] Harper & Row, Publ'rs, Inc. v. Nation Enters., 471 U.S. 539, 564–65 (1985).

[28] Sheldon v. Metro-Goldwyn Pictures Corp., 81 F.2d 49, 54-56 (2d Cir. 1936).

[29] Bridgeport Music, Inc. v. Dimension Films, 410 F.3d 792, 801–02 (6th Cir. 2005).

[30] 17 U.S.C. § 106(2).

[31] Castle Rock Entm't, Inc. v. Carol Publ'g Grp., Inc., 150 F.3d 132, 140–42, 145–46 (2d Cir. 1998).

[32] *See, e.g.*, Case C-5/08, Infopaq Int'l A/S v. Danske Dagblades Forening, 2009 E.C.R. I-6569, ¶ 48 (read with ¶ 37), ECLI:EU:C:2009:465 (holding that reproduction of an 11-word extract could infringe if it contained an element expressive of the author's own intellectual creation);

Jurisprudence in music and visual art illustrates this spectrum of analysis, all predicated on the court's ability to compare the works directly. While *Bridgeport* adopted an almost absolute rule for audible sound recording samples, the Ninth Circuit, in *VMG Salsoul, LLC v. Ciccone*, carved out a *de minimis* exception where a 0.23-second horn hit was deemed undetectable by ordinary listeners and thus non-infringing.[33] Visual art cases show a similar reliance on direct comparison: Jeff Koons's near-verbatim sculptural copy of a photograph was condemned in *Rogers v. Koons*,[34] yet many of Richard Prince's collage canvases, which recast Rastafarian portraits into a "fundamentally different aesthetic," were absolved as transformative in *Cariou v. Prince*.[35]

When substantial similarity is undeniable, or a work is clearly derivative, the doctrine of *transformative fair use*, articulated in *Campbell v. Acuff-Rose Music, Inc.*, provides a crucial escape valve.[36] Copying may be excused if the new work adds "new expression, meaning, or message" for a distinct purpose, thereby transforming the original rather than merely superseding it. This doctrine has been pivotal in cases involving information-extraction technologies, most notably in *Authors Guild v. Google, Inc.*, which permitted full-text scanning of books for a search index that displayed only non-substitutive snippets,[37] and in software reverse-engineering cases like *Sega Enterprises Ltd. v. Accolade, Inc.* and *Sony Computer Entertainment, Inc. v. Connectix Corp.*, where intermediate copying was allowed to create new, non-infringing interoperable products.[38] However, even these transformative use victories often hinged on a concrete comparison and the fact that the infringing copies were either invisible to the public or the outputs were demonstrably non-competitive. The Supreme Court's recent decision in *Andy Warhol Foundation for the Visual Arts, Inc. v. Goldsmith* has further tightened this lens, stressing that adding "new meaning or message" is insufficient for fair use if the secondary use ultimately serves the *same* commercial purpose as, and competes with, the original.[39]

Crucially, in these and similar instances, the court possessed *both* works and could meticulously assess whether and how protected elements traveled from one to the other. Notably absent from the case-law architecture is any sustained inquiry into the *process* by which the defendant's work was produced, *unless that process itself involved an actionable reproduction*.

---

Designers Guild Ltd. v. Russell Williams (Textiles) Ltd., [2000] UKHL 58, [11]–[25] (discussing substantial part in the context of artistic works).

[33] VMG Salsoul, LLC v. Ciccone, 824 F.3d 871, 877–78 (9th Cir. 2016).

[34] Rogers v. Koons, 960 F.2d 301, 307–09 (2d Cir. 1992).

[35] Cariou v. Prince, 714 F.3d 694, 708–09 (2d Cir. 2013).

[36] Campbell v. Acuff-Rose Music, Inc., 510 U.S. 569, 579 (1994).

[37] Authors Guild v. Google, Inc., 804 F.3d 202, 214–19, 225 (2d Cir. 2015).

[38] Sega Enters. Ltd. v. Accolade, Inc., 977 F.2d 1510, 1522–28 (9th Cir. 1992); Sony Computer Entm't, Inc. v. Connectix Corp., 203 F.3d 596, 602–07 (9th Cir. 2000).

[39] Andy Warhol Found. for the Visual Arts, Inc. v. Goldsmith, 143 S. Ct. 1258, 1282-83 (2023).

Whether the copyist used tracing paper, a musical sampler, or a digital scanner mattered primarily insofar as protectable expression ultimately surfaced in the output. Indeed, as cases like *Sega* and *Google Books* illustrate, even intermediate steps involving extensive, prima facie infringing copying could be excused if the final product was sufficiently transformed and served a legitimate, non-infringing purpose.[40] Liability, in essence, has consistently traced the path of expression, not the intricacies of the creative or reproductive process itself.[41]

Taken together, these disparate doctrines form an intricate "evidence machine" that functions only when courts can place the original work and the allegedly infringing creation on the same analytic table. For all its complexity, the machine runs on a single fuel: *visible, comparable artifacts*. The next Part shows how the opaque, recursive pipelines of modern AI siphon off that fuel—stalling the evidence machine and forcing a search for a new anchor for liability.

# II THE AI OUROBOROS PROBLEM: TOWARD AN AI-FOPT DOCTRINE

## A. Copyright Laundering and the Collapse of Visibility

The *presumption of visibility*—the bedrock of infringement analysis—falters when confronted with the "AI Ouroboros": a multi-generational training pipeline that uses recursive synthetic data to systematically launder copyright infringement. Outputs from AI models—across successive model generations—can be systematically "scrubbed"[42] of overtly infringing content, creating a chain of training datasets, each derived from the outputs of its predecessor. This process progressively obscures the link between a late-generation model's outputs and the original copyrighted material, such that even if a final model's capabilities derive from the value inherent in original protected content, its outputs can be practically untraceable to any specific copyrighted source.

[40] Sega, 977 F.2d at 1522–28; Sony, 203 F.3d at 602–07; Authors Guild, 804 F.3d at 225.

[41] *See* Kelly v. Arriba Soft Corp., 336 F.3d 811, 822 (9th Cir. 2003) (finding the creation and display of thumbnail images in a search engine to be a transformative fair use); Perfect 10, Inc. v. Amazon.com, Inc., 508 F.3d 1146, 1164 (9th Cir. 2007) (affirming the transformative nature of image search).

[42] *"Scrubbed"* here refers to various potential techniques for processing AI-generated outputs to reduce or eliminate traces of specific training data. This could involve, for example, filtering mechanisms to detect and remove content that directly matches known copyrighted material, or more sophisticated methods where an intermediate AI model might be used to paraphrase, rephrase, or otherwise transform outputs to obscure similarity to original sources while retaining underlying concepts or styles. *See, e.g.*, Nikhil Vyas et al., *On Provable Copyright Protection for Generative Models*, Proc. 40th Int'l Conf. on Machine Learning (ICML), at 35277-35299 (2023) (proposing algorithms that modify generative model training to ensure outputs probabilistically diverge from copyrighted training data); *see also* Pratyush Maini et al., *TOFU: A Task of Fictitious Unlearning for LLMs*, arXiv:2401.06121 (2024), https://arxiv.org/abs/2401.06121 [https://perma.cc/L2AR-LMY3] (evaluating "unlearning" techniques designed to remove specific data from trained models); *cf.* Debalina Padariya et al., *Privacy-Preserving Generative Models: A Comprehensive Survey*, arXiv:2502.03668 (Feb. 6, 2025),

The evidentiary impasse is compounded by the sheer complexity of modern AI development. Large-scale, multi-generational AI systems are often developed across multiple teams and organizations, each potentially contributing distinct components or sub-models.[43] These elements are frequently integrated into massive, complex architectures, further diffusing the ability to pinpoint precisely where and how copyrighted material was initially incorporated or how its influence propagated.[44] As a result, satisfying the traditional evidentiary requirements of demonstrating direct access to an original work and establishing substantial similarity in a final output can become an almost insurmountable task.

## B. The Inadequacy of Existing Doctrinal Tools

While the law possesses a range of tools to address hard-to-prove copying, none proves adequate to police these unique evidentiary challenges. When evaluated against the standard of doctrinal workability—that is, a judge-made rule that fits within existing judicial powers, restores a realistic evidentiary pathway, and is cabined by familiar limits—each of the plausible alternatives falls short. They either (i) presuppose the very evidence that copyright laundering is designed to erase; (ii) require a controversial expansion of substantive rights; (iii) are too episodic to offer a systemic solution; or (iv) depend on legislative action beyond the judiciary's power to implement.

---

https://arxiv.org/abs/2502.03668 [https://perma.cc/X42E-JS7Y] (surveying privacy-preserving techniques that function analogously for copyright protection).

[43] The collaborative nature of large-scale AI development is evident in major open-source initiatives where multiple institutions and research groups contribute to building and training complex models. *See, e.g.*, BIGSCIENCE WORKSHOP, *BLOOM: A 176B-Parameter Open-Access Multilingual Language Model*, ARXIV:2211.05100 (Nov. 9, 2022), https://arxiv.org/abs/2211.05100 [https://perma.cc/8N42-6SD3] (describing the creation of a large language model through a broad international research collaboration, exemplifying distributed development).

[44] The development of "foundation models," which underpin many contemporary AI applications, typically involves substantial, often collaborative, efforts across various stages, from data curation to pre-training and fine-tuning, reflecting a complex, multi-faceted development pipeline that inherently complicates tracing the influence of specific training data. *See, e.g.*, Rishi Bommasani et al., *On the Opportunities and Risks of Foundation Models*, ARXIV:2108.07258 (Aug. 16, 2021), https://arxiv.org/abs/2108.07258 [https://perma.cc/3U2V-6DGV] (providing a comprehensive overview of foundation models, their ecosystem, and the significant resources and often distributed efforts involved in their creation).

The most obvious tools, the output-focused tests of access and substantial similarity, are the first to fail. These doctrines are fundamentally premised on the "presumption of visibility" established in Part I; they function only when there are stable, comparable artifacts to place side-by-side.[45] The entire purpose of copyright laundering is to systematically break this visibility, ensuring that any infringing expression is so diffused and abstracted that it no longer registers as substantially similar. Plaintiffs are thus left with a proof problem that is not incidental but engineered.

Nor can this failure be remedied by simply adjusting the sensitivity of the similarity test. Tightening the standard to require "striking similarity"—which applies only when similarities are so strong as to preclude the possibility of independent creation—is of little help.[46] This high bar will rarely be met in the context of laundered infringement, as modern models are explicitly trained to disperse expression and avoid the kind of stark, congruent overlap the doctrine requires, even while the influence of original works persists as extractable memorization.[47] Conversely, loosening the standard to cover "total concept and feel" becomes hopelessly speculative when a model is trained on a corpus as vast as the internet, dramatically increasing the risk of punishing mere coincidence.[48]

Expanded liability theories prove equally unworkable. Secondary liability doctrines, such as contributory and vicarious infringement, are powerful but require an identifiable act of direct infringement to which liability can attach.[49] In the laundering context, the primary wrong is the *upstream* act of unlawful training, while the direct infringement, if any, is committed by a diffuse set of *downstream* end-users generating facially non-infringing outputs.

Alternatively, attempting to expand copyright's scope to protect an author's "style" is doctrinally untenable. The courts have consistently afforded only "thin" protection to an author's style,[50] and such a move would require courts to abandon the idea/expression dichotomy and disregard the Supreme Court's instruction that copyright does not protect processes, systems, or unoriginal compilations of fact.[51]

---

[45] *See, e.g.*, Arnstein v. Porter, 154 F.2d 464, 468-69 (2d Cir. 1946).

[46] *See* Baxter v. MCA, Inc., 812 F.2d 421, 423 (9th Cir. 1987); Selle v. Gibb, 741 F.2d 896, 903-06 (7th Cir. 1984).

[47] *Cf.* Aneesh Pappu et al., *Measuring Memorization in RLHF for Code Completion*, ARXIV:2406.11715 (June 17, 2024), https://arxiv.org/abs/2406.11715 [https://perma.cc/Q4SH-TFVY] (finding RLHF reduces but does not eliminate memorization); Milad Nasr et al., *Scalable Extraction of Training Data from (Production) Language Models*, ARXIV:2311.17035 (Nov. 28, 2023), https://arxiv.org/abs/2311.17035 [https://perma.cc/SBP4-NZH2] (demonstrating that training data remains extractable even from aligned production models).

[48] *Cf.* Nichols v. Universal Pictures Corp., 45 F.2d 119, 121 (2d Cir. 1930); *see also* Peter Pan Fabrics, Inc. v. Martin Weiner Corp., 274 F.2d 487, 489 (2d Cir. 1960); Roth Greeting Cards v. United Card Co., 429 F.2d 1106, 1110 (9th Cir. 1970) (coining "total concept and feel").

[49] *See, e.g.,* MGM Studios, Inc. v. Grokster, Ltd., 545 U.S. 913, 929-35 (2005); A&M Records, Inc. v. Napster, Inc., 239 F.3d 1004, 1019–25 (9th Cir. 2001).

Other process-focused and equitable tools are too episodic to offer a systemic solution. Spoliation sanctions, for instance, are designed to punish the destruction of evidence that a party had a duty to preserve, a duty that is often triggered only by reasonably foreseeable litigation.[52] Copyright laundering, however, involves an *ex ante design choice* that avoids creating stable evidentiary artifacts in the first place. Likewise, while unjust enrichment can disgorge profits from a proven wrong, it does not provide the underlying liability rule for determining *whether* a wrong has occurred.

Finally, while legislative reforms are valuable, they are not doctrinal solutions that courts can apply today. Transparency mandates, such as the EU AI Act's requirement for a "sufficiently detailed summary" of training data, are a crucial step forward.[53] However, a narrative summary still leaves the plaintiff with the impossible burden of tracing provenance. Transparency is an essential *input* for a legal test; it is not the test itself.

## C. The Search for a New Anchor: The FOPT Analogy

With existing tools proving inadequate, the task is to find a new, judicially administrable anchor for liability—one that can trace the taint of infringement through the technological chain of custody. The FOPT doctrine—originally developed in the distinct legal domain of U.S. criminal procedure—offers that analogical anchor. In its home context, FOPT dictates that evidence obtained as a result of an illegal search or seizure is inadmissible in court; the initial illegality (the "poisonous tree") taints any subsequently derived evidence (the "fruit").[54] Its core tenet is institutional: the law should not permit the state to benefit from its own unlawful actions, thereby deterring such misconduct at its source and preserving judicial integrity. Applied by analogy to AI systems, if the unauthorized copying of copyrighted material renders the training of an initial AI model ($AI_1$) an act of infringement (a "poisonous tree," as potentially established in cases like *Thomson Reuters Enter. Ctr. GmbH v. Ross Intelligence Inc.*),[55] then subsequent AI models ($AI_2$, $AI_3$, etc.) trained primarily on synthetic datasets derived from $AI_1$'s outputs or formed by distillation from its weights—that is, the "tree's" tainted "fruits"—would presumptively bear a taint.

---

[50] *See* Satava v. Lowry, 323 F.3d 805, 812–13 (9th Cir. 2003) (holding that realistic glass-in-glass jellyfish sculptures composed of unprotectable ideas and standard elements receive only "thin" copyright and cannot monopolize those elements).

[51] *See* Feist Publ'ns, Inc. v. Rural Tel. Serv. Co., 499 U.S. 340, 349–50 (1991); 17 U.S.C. § 102(b) (2018).

[52] *See, e.g.*, Zubulake v. UBS Warburg LLC (Zubulake IV), 220 F.R.D. 212, 216-17 (S.D.N.Y. 2003); Zubulake v. UBS Warburg LLC (Zubulake V), 229 F.R.D. 422, 436–37 (S.D.N.Y. 2004); FED. R. CIV. P. 37(e).

[53] *See* Artificial Intelligence Act, *supra* note 5, art. 53(1)(d) (requiring providers of GPAI models to "draw up and make publicly available a sufficiently detailed summary about the content used for training"); *id.* Recital 107 (explaining that providers should list the main data sources and may provide the required summary "in narrative form").

[54] *See, e.g.,* Silverthorne Lumber Co. v. United States, 251 U.S. 385, 392 (1920); Nardone v. United States, 308 U.S. 338, 341–42 (1939); Wong Sun v. United States, 371 U.S. 471, 488 (1963).

While copyright law does not currently incorporate a direct FOPT rule for infringement, generally focusing on whether the *end product itself* infringes an exclusive right rather than automatically tainting a non-infringing downstream product solely because of an illicit step in its creation, scholars have explored conditions under which such reasoning might be justified. Perhaps most crucially, Lemley, while cautioning against the broad, uncritical importation of FOPT into intellectual property doctrine due to risks of overreach,[56] proposes a test for human-driven creation hinging on (i) the wrongdoer's mental state, (ii) the likelihood that the infringement will be detected, and (iii) the relative contribution of the final, non-infringing product.

Recursive AI training, however, makes Lemley's test impracticable. While this test represents a valuable endeavor to bring consistency to FOPT considerations within IP, particularly for scenarios involving human-driven creation or more transparent derivative chains, it becomes largely unworkable when confronted with the realities of modern AI development.

*First*, the "mental state of the infringer" inquiry becomes elusive. While human intent is central to the initial decision to scrape data for a foundational model ($AI_1$), ascribing a legally cognizable "mental state" to the subsequent operations of the model itself, or to the diffuse, often automated processes that use $AI_1$'s outputs to train $AI_2$ and beyond, becomes problematic. The human decision-making behind the propagation of data through these pipelines can be so distributed and algorithmically mediated as to render a unified "wrongdoer's intent" for each stage practically infeasible to infer ex post.

*Second*, the "ease of enforcing the IP right against the actually infringing products" or the likelihood of detection shifts dramatically. While the initial ingestion by $AI_1$ might be identifiable as infringing, the subsequent outputs of $AI_1$, and the models ($AI_2$, $AI_3$…) trained on them, may not overtly reproduce specific copyrighted expression in a way that traditional access and substantial similarity tests can readily capture—the likelihood of detecting a specific, actionable infringement in the output of a late-generation model trained on billions of data points, whose outputs are stochastic and potentially infinite, diminishes significantly.

*Third*, assessing the relative contribution of any single illicitly used input to $AI_1$ versus the value added by $AI_1$'s own transformative processing, or the contribution of $AI_1$'s outputs to the capabilities of a general-purpose $AI_2$ (and so on), becomes an exercise in deep speculation, especially at the point of creation and deployment when the vast range of downstream applications is unknown.

---

[55] Thomson Reuters Enter. Ctr. GmbH v. Ross Intel. Inc., 765 F. Supp. 3d 382, 391 (D. Del. 2025) (Order Granting Partial Summary Judgment).

[56] Lemley, *supra* note 15, at 264–65.

The AI Ouroboros thus presents a scenario where the iterative abstraction, transformation, and commingling of sources across opaque technological layers demands an adapted FOPT logic. Rather than re-evaluating Lemley's challenging factors at each derivative stage within an AI pipeline, this Article argues for an AI-FOPT standard keyed to the adjudicated status of the foundational model: If the training of an initial AI model ($AI_1$) is adjudged infringing (a "poisonous tree," due to unauthorized copying not excused by fair use), then the datasets and subsequent AI models ($AI_2$, $AI_3$, etc.) *principally derived* from $AI_1$'s outputs (e.g., where $AI_1$'s outputs form a substantial basis for the weights, architecture, or training data of subsequent models) inherit a *rebuttable presumption* of taint. Once triggered, this presumption applies *categorically*, shifting the evidentiary burden to downstream developers to demonstrate a clean lineage, without requiring a fresh application of Lemley's three factors to each subsequent, synthetically generated iteration.

The proposed AI-FOPT doctrine is a narrow, lineage-focused supplement, and not a wholesale replacement of traditional copyright analysis or even of Lemley's framework where his factors remain tractable. Existing frameworks, including access and substantial similarity tests, continue to govern human-authored works or first-generation AI outputs where direct comparison is feasible. The determination of whether $AI_1$ itself constitutes an infringing "poisonous tree" relies entirely on established legal principles, where factors like intent and the nature of the use remain pertinent to assessing fair use. The AI-FOPT presumption activates only *after* a foundational infringement has been adjudicated. Its role is to address the evidentiary void created by recursive AI training and address the systemic risk of *copyright laundering* created when multi-generational training pipelines render traditional proof of infringement for downstream creations practically impossible. Human creativity and AI systems whose lineages do not pass through an adjudicatedly infringing model remain governed exclusively by existing doctrines.

Such an AI-FOPT standard, while novel in its direct application of FOPT to copyright,[57] coheres with familiar legal principles. For instance, the *Bridgeport* court's strict "no sampling"

---

[57] Historically, there has been a general absence of FOPT in copyright law and preparatory infringement can lead to non-infringing subsequent products. *See, e.g.*, Matthew Sag, *Fairness and Fair Use in Generative AI*, 92 FORDHAM L. REV. 1887, 1918 n.207 (2024) (discussing the non-recognition of FOPT in copyright); *see also* Kepner-Tregoe, Inc. v. Leadership Software, Inc., 12 F.3d 527, 538 (5th Cir. 1994), and Lemley, *supra* note 15, at 246, 260-261. However, the unique challenges posed by generative AI are already prompting plaintiffs in several high-profile cases to seek remedies that directly reflect a lineage-based taint theory, asking courts to order the destruction of the models themselves. *See, e.g.*, Complaint, Prayer for Relief K, *Getty Images (US), Inc. v. Stability AI, Inc.*, No. 1:23-cv-00135 (D. Del. filed Feb. 3, 2023) ("Ordering the destruction of all versions of Stable Diffusion trained using Getty Images' content without permission"); Complaint, Prayer for Relief ¶¶ f, h, *Ziff Davis, Inc. v. OpenAI, Inc.*, No. 1:25-cv-00501 (D. Del. filed Apr. 24, 2025) (requesting, pursuant to 17 U.S.C. § 503(b), the "destruction of all OpenAI LLMs, products, systems, and training sets incorporating the Registered Works"); Complaint, Prayer for Relief ¶ 3, *N.Y. Times Co. v. Microsoft Corp.*, No. 1:23-cv-11195

rule for sound recordings effectively attaches liability to the *act of taking* an identifiable fragment, regardless of downstream transformation if that fragment remains recognizable. Copyright law routinely bars "bootleg" chains: selling DVDs pressed from an illicit master recording is infringement, even if the retail disc itself bears no immediately visible link to the original unauthorized rip. Remedial analogues abound, such as court orders for the destruction of infringing molds (as in *Rogers v. Koons*), the recall of infringing CDs, or the exclusion of misappropriated code in software cases, all reflecting a judicial willingness to cut off downstream exploitation once a root violation has been established. What unites these precedents is the refusal to allow subsequent transformation to sanitize an upstream wrong—a principle this Article adapts to the case of recursive synthetic data training.

The next Part details the operational mechanics of the proposed doctrine—how the taint of an initial infringement would attach, how the presumption could be rebutted, and what remedial consequences might follow.

# III. OPERATIONALIZING AI-FOPT: A JUDICIALLY ADMINISTRABLE FRAMEWORK

This Part translates the "fruit of the poisonous tree" analogy into a legally workable doctrine. The proposed AI-FOPT standard is not a standalone cause of action but an evidentiary rule designed to restore the bargain of proof that copyright laundering seeks to break. Grounded in familiar principles from copyright, trade secret, and procedural law, it shifts the burden of production to the party with unique and low-cost access to the dispositive evidence of a model's lineage—the developer—once a plaintiff makes a threshold showing of foundational infringement and downstream derivation.[58] The framework proceeds in five steps—trigger, derivation, presumption, rebuttal, and remedies—which the subsections below in turn develop.

**The AI-FOPT Operational Test**

1. **Trigger (Poisonous Tree):** A court *adjudges* that a foundational model (poisonous tree) was trained via unauthorized copying not excused by fair use or other defenses.

---

(S.D.N.Y. filed Dec. 27, 2023) (requesting "destruction under 17 U.S.C. § 503(b) of all GPT or other LLM models and training sets that incorporate Times Works"). *See also* Daniel Wilf-Townsend, *The Deletion Remedy*, 103 N.C. L. REV. 1809 (2025) (analyzing these and other requests for model deletion as a novel remedy in copyright law).

[58] The principle that evidentiary burdens may be adjusted to account for one party's superior access to proof is a cornerstone of civil procedure, particularly in cases involving complex technology or corporate defendants. *See, e.g.*, Zubulake v. UBS Warburg LLC (Zubulake IV), 220 F.R.D. 212, 220 (S.D.N.Y. 2003); *see also* 2 CHARLES T. MCCORMICK, MCCORMICK ON EVIDENCE § 337 (8th ed. 2020) (discussing the allocation of burdens of proof based on factors including "the defendant's superior access to the proof").

2. **Derivation (Principally Derived):** The plaintiff makes a prima facie showing that a challenged model is *principally derived* from the poisonous tree's (or its successor models') outputs or distilled weights (e.g., initialized, distilled, or merged from the poisonous tree or its tainted successors, materially reliant on synthetic data from the tainted lineage; assessed at the component-level for MoE).

3. **Presumption & Burdens:** A rebuttable presumption of taint attaches. The *burden of production* shifts under Fed. R. Evid. 301. Given asymmetrical access to provenance, courts may treat rebuttal as an *affirmative defense*, placing the *burden of persuasion* on the developer.

4. **Rebuttal Paths:** By a preponderance, the developer shows either:

    a. **Clean Lineage** (auditable, license-cleared, independent training provenance), or

    b. **Purged Taint** (curative rebuild or effective unlearning), verified by pre-registered, performance-based audits *admissible under* Rule 702/Daubert.

2. **Remedies:** If unrebutted, courts apply a *calibrated ladder*: targeted/component-level injunctions (including head-start relief), ongoing royalties or profits (17 U.S.C. § 504), and in exceptional cases impoundment/destruction (17 U.S.C. § 503), consistent with *eBay*.

## A. The Trigger: Adjudicated Illegality and the "Poisonous Tree"

AI-FOPT activates only upon the identification of a "poisonous tree": an AI model ($AI_1$) whose training has been *adjudged* infringing. The predicate is a judicial finding that $AI_1$ was trained on copyrighted works without authorization and that the copying is *not* excused by fair use or another defense. This requirement anchors the doctrine in established copyright principles and prevents the creation of a new, process-based right, leaving the fair-use inquiry at ingestion as the dispositive gatekeeper. *Thomson Reuters Enter. Ctr. GmbH v. ROSS Intelligence Inc.* provides a paradigmatic trigger: the court held that wholesale ingestion of Westlaw's copyrighted headnotes to build a competing research tool failed fair-use scrutiny,[59] rendering the model a "poisonous tree."

Conditioning the trigger on a prior adjudication promotes procedural efficiency. Under the principle of non-mutual offensive issue preclusion, a final judgment that a specific foundational model was unlawfully trained can be invoked by subsequent plaintiffs *against the same developer (or those in privity)*, subject to the fairness factors articulated in *Parklane*

[59] Thomson Reuters Enter. Ctr. GmbH v. Ross Intel. Inc., 765 F. Supp. 3d 382 (D. Del. 2025) (Order Granting Partial Summary Judgment).

*Hosiery Co. v. Shore*.[60] This allows courts to establish the "poisonous tree" element without relitigating the complex facts of the foundational model's training, conserving judicial resources and ensuring consistent treatment of an infringing model at its root.

Existing and emerging transparency mandates can serve as crucial *discovery footholds* for establishing this trigger. In the European Union, for instance, providers of general-purpose AI models must include a "sufficiently detailed summary" of training content in their technical documentation.[61] In the United States, the Copyright Office has signaled the importance of provenance even while stopping short of recommending a formal mandate.[62] While these instruments do not themselves decide infringement, they provide a procedural on-ramp for plaintiffs to gather the evidence necessary to prove that a model's training corpus contained unauthorized works, thereby helping to establish the "poisonous tree" predicate for the AI-FOPT framework. Indeed, recent litigation demonstrates that courts can compel the disclosure of such provenance records, which can prove decisive.[63]

## B. The Scope of Taint: Defining "Principally Derived"

Once $AI_1$ is established as a poisonous tree, its taint extends to any subsequent model or dataset that is "principally derived" from it. This standard captures the concept of *material dependence*, assessed through a fact-intensive inquiry that considers both quantitative and qualitative factors.

---

[60] *See* Parklane Hosiery Co. v. Shore, 439 U.S. 322, 331–33 (1979) (permitting a plaintiff to use a prior judgment offensively to prevent a defendant from relitigating an issue that the defendant had previously lost in an earlier action).

[61] Artificial Intelligence Act, *supra* note 5, art. 53(1)(d).

[62] U.S. COPYRIGHT OFFICE, *supra* note 5, at 100-01, 106 (summarizing commenters' proposals emphasizing transparency/identification of training data and recommending that the licensing market be allowed to develop without new government mandates).

[63] The practical importance of such discovery was highlighted in the recent *Anthropic* copyright litigation. After discovery, the court issued a key ruling on fair use that relied on detailed evidence of the defendant's training corpora. *See* Bartz v. Anthropic PBC, 787 F. Supp. 3d 1007, 1030-31 (N.D. Cal. 2025) (Order on Fair Use); *id.* at 30–31 (denying summary judgment on fair use for the retained "pirated library copies," distinguishing them from the potentially transformative act of training). That order was followed by a proposed $1.5 billion class action settlement. The court then postponed preliminary approval, ordering the parties to submit a final, detailed "Works List." *See* Minute Entry for Proceedings Held on Sept. 8, 2025 (Doc. 371), *Bartz v. Anthropic PBC*, No. 3:24-cv-05417-WHA (N.D. Cal. Sept. 8, 2025). In a subsequent order, the court probed the specifics of the settlement, questioning the "documentary proof" required for works on the "Works List" and the lack of any injunctive relief to "delete any copies downloaded from Books3." *See* Order, Questions for Preliminary Approval Hearing on September 25, at 2–3 (Q11, Q15) (Doc. 375); *see also* Blake Brittain, *Anthropic's $1.5 billion copyright settlement faces judge's scrutiny*, REUTERS (Sept. 9, 2025), https://www.reuters.com/legal/government/anthropics-15-billion-copyright-settlement-faces-judges-scrutiny-2025-09-09/ [https://perma.cc/A3JV-UKEC] (reporting on the hearing).

To make it judicially administrable, courts should look for several strong, non-exhaustive indicia of derivation:

- **Initialization, Distillation, or Merging.** A downstream model ($AI_2$) is principally derived if it is initialized from $AI_1$'s weights, created through a distillation process where $AI_1$ acts as the "teacher" model, or formed by merging $AI_1$'s parameters with another model. In such cases, the entire parameter space of $AI_2$ inherits the learned representations—and thus the taint—of its predecessor.

- **Substantial Synthetic Data Reliance.** Derivation is also established where a material portion of $AI_2$'s training or fine-tuning corpus consists of synthetic data generated by $AI_1$. This directly addresses the core laundering mechanism, where the outputs of an infringing model are repackaged as a seemingly "clean" dataset.

- **Component-Level Derivation.** In modular architectures, such as Mixture-of-Experts (MoE) systems, taint may attach at a component level. If specific experts or modules within a larger model were trained on $AI_1$'s outputs or distilled weights, those components are deemed principally derived, allowing for more targeted discovery and remedies.

This standard is carefully cabined to avoid chilling legitimate innovation. For example, fine-tuning a lawfully trained, public-domain base model exclusively on license-cleared, proprietary corporate data would ordinarily not create a "principally derived" model, as its new capabilities do not materially depend on the outputs or distilled weights of an infringing predecessor.

This approach borrows from established copyright doctrines. The "material dependence" standard parallels the "substantial part" analysis used to assess derivative works, which evaluates not just the quantity but the qualitative importance of the appropriated expression.[64] Similarly, the ability to analyze modular components separately tracks the Abstraction-Filtration-Comparison test used in software copyright cases, where courts methodically dissect complex programs to isolate infringing elements from non-infringing ones.[65]

## C. The Presumption: Shifting the Burdens of Proof

Upon a plaintiff's prima facie showing of both an adjudicated "poisonous tree" and that the defendant's model is principally derived from it, a rebuttable presumption of taint attaches to the defendant's model. This presumption serves a critical function: it reallocates the evidentiary burdens to align with the realities of information access in opaque AI development. Under

---

[64] *See, e.g.,* Castle Rock Entm't, Inc. v. Carol Publ'g Grp., Inc., 150 F.3d 132, 140–42 (2d Cir. 1998) (discussing the different tests in comparing an original and derivative works).

[65] *See* Computer Assocs. Int'l, Inc. v. Altai, Inc., 982 F.2d 693, 706–11 (2d Cir. 1992).

Federal Rule of Evidence 301, this presumption shifts the *burden of production* to the developer to come forward with evidence of a clean lineage or a curative rebuild.[66]

However, given the profound information asymmetries inherent in multi-generational AI pipelines, where evidence of provenance is uniquely and exclusively held by the developer, courts should, in equity, treat the rebuttal pathways of "clean lineage" or "purged taint" as affirmative defenses. This would place not only the burden of production but also the *burden of persuasion* on the developer to prove, by a preponderance of the evidence, that its model is not tainted. This allocation is justified by the prohibitive, if not impossible, task a plaintiff would otherwise face in trying to prove the negative—that no clean, independent source exists. This approach aligns with principles underlying secondary copyright liability, where a party with knowledge and material contribution to infringement bears a burden to demonstrate its non-involvement,[67] and it tracks the logic of the spoliation doctrine, where the failure to preserve or produce evidence to which a party has unique access can lead to adverse inferences or the shifting of evidentiary burdens.[68]

A familiar statutory analogue in patent law underscores this allocation. In process-patent cases, where an infringing manufacturing process is often as opaque as an AI training pipeline, Congress has determined that a simple shift in the burden of production is insufficient. Under 35 U.S.C. § 295, once a plaintiff makes a threshold showing, the *burden of persuasion* shifts to the accused infringer to prove that its product was made by a non-infringing process.[69] The same logic applies with equal, if not greater, force here: model training is a non-public process whose provenance records are uniquely within the developer's control, justifying a similar shift in the ultimate burden of proof.

## D. Rebuttal Pathways: Proving a Clean or Cured Lineage

A developer can rebut the presumption of taint by proving, by a preponderance of the evidence, that its model follows one of two paths: it either originates from an entirely independent source or has been effectively purged of the initial infringement. These pathways are analogous to the "independent source" and "attenuation" exceptions to the exclusionary rule in criminal procedure, adapted to the context of generative AI.

1. **Independent Source (Clean Lineage).** The most direct rebuttal is to demonstrate a verifiably independent training history for the allegedly tainted model ($AI_2$). This requires the production of auditable records—a "Provenance Packet"—that may include cryptographically hashed manifests of all training and fine-tuning datasets, licenses for any third-party data, and checkpoint lineage graphs that affirmatively show $AI_2$ was not initialized from or otherwise derived from the weights of the poisonous model ($AI_1$). This pathway is functionally equivalent to the "clean room" development process, which has been sanctioned by courts in software copyright cases as a legitimate method for creating a non-infringing product without reference to protected source code.[70]

---

[66] FED. R. EVID. 301 ("[T]he party against whom a presumption is directed has the burden of producing evidence to rebut the presumption.").

[67] *See* Gershwin Publ'g Corp. v. Columbia Artists Mgmt., Inc., 443 F.2d 1159, 1162 (2d Cir. 1971) (defining contributory infringement as occurring when one, "with knowledge of the infringing activity, induces, causes or materially contributes to the infringing conduct of another").

[68] *See, e.g.,* Zubulake v. UBS Warburg LLC (Zubulake V), 229 F.R.D. 422, 436–37 (S.D.N.Y. 2004).

[69] 35 U.S.C. § 295.

2. **Purged Taint (Curative Rebuild or Unlearning).** Alternatively, a developer can prove that the inherited taint has been affirmatively and effectively purged. This requires more than superficial "scrubbing" or cosmetic alterations of $AI_1$'s outputs, which would be insufficient to cure the taint just as minor changes to a copied photograph were insufficient in *Rogers v. Koons*.[71] Instead, the developer must demonstrate a genuine cure, either through a complete *curative rebuild*—retraining the model from scratch on a verifiably clean corpus—or through the use of robust and verifiable *machine unlearning* techniques that excise the influence of $AI_1$.

The effectiveness of any claimed purge must be established through rigorous, independent, performance-based audits, with the methodology and results methodology and results subject to the *Daubert* standard for expert evidence.[72] While courts may consider empirical evidence from technical probes such as model-inversion or membership-inference attacks, such evidence should be treated as *corroborative, not dispositive*, given their known error rates. To ensure reliability, any such technical evidence must be supported by pre-registered, reproducible protocols and should be triangulated with documentary evidence of provenance, such as the logs and manifests contained in the Provenance Packet.

## E. Remedies: A Calibrated Ladder

If the developer fails to rebut the presumption of taint, a court must fashion an appropriate remedy. Rather than applying a categorical rule, courts should tailor relief using a calibrated ladder of remedies, guided by the traditional four-factor test for equitable relief articulated in *eBay Inc. v. MercExchange, L.L.C.*[73] This approach ensures that the remedy is proportional to the harm and the nature of the infringement. The ladder of potential remedies includes:

---

[70] *See* Sony Computer Entm't, Inc. v. Connectix Corp., 203 F.3d 596, 602–07 (9th Cir. 2000); Sega Enters. Ltd. v. Accolade, Inc., 977 F.2d 1510, 1522–28 (9th Cir. 1992).

[71] *See* Rogers v. Koons, 960 F.2d 301, 307–09 (2d Cir. 1992).

[72] Daubert v. Merrell Dow Pharms., Inc., 509 U.S. 579, 589–95 (1993) (establishing the standard for the admissibility of expert scientific testimony in federal court).

[73] eBay Inc. v. MercExchange, L.L.C., 547 U.S. 388, 391 (2006) (holding that a plaintiff seeking a permanent injunction must demonstrate: (1) that it has suffered an irreparable injury; (2) that remedies available at law, such as monetary damages, are inadequate to compensate for that injury; (3) that, considering the balance of hardships between the plaintiff and defendant, a

- **Targeted Injunctive Relief.** A court could issue an injunction that is narrowly tailored to the scope of the taint. Consistent with the principle of proportionality, where taint is confined to specific components in a modular model (such as certain experts in a Mixture-of-Experts system), courts should target relief to specific components where feasible—for example, by disabling the infringing modules pending a curative rebuild. Analogizing to trade secret law, which aims to neutralize the competitive advantage gained from misappropriation, a court could also impose a "head-start" injunction, barring the model's use for a period equivalent to the time it would have taken to develop it lawfully.[74]

- **Monetary Relief.** Where a permanent injunction is disproportionate, a court might order an ongoing royalty for future use of the tainted model. For past infringement, a plaintiff would be entitled to actual damages and any of the infringer's profits attributable to the infringement. Crucially, under the Copyright Act, the defendant bears the burden of apportioning its profits between infringing and non-infringing factors.[75]

- **Impoundment or Destruction.** In exceptional cases, such as those involving willful and continued infringement or where the taint is so inextricably woven into the model that it cannot be purged, a court may order the impoundment or destruction of the infringing model and its derivative artifacts, as authorized by the Copyright Act.[76]

Any request for preliminary relief, such as an order halting the deployment of a model pending a final decision, would be governed by the familiar standard set forth in *Winter v. Natural Resources Defense Council, Inc*.[77]

## F. Evidentiary and Procedural Mechanics

To be judicially administrable, the AI-FOPT framework must be supported by a structured approach to discovery that balances a plaintiff's need for evidence against a developer's legitimate interest in protecting trade secrets. Courts can achieve this balance by adopting a two-

---

remedy in equity is warranted; and (4) that the public interest would not be disserved by a permanent injunction).

[74] *See, e.g.,* Winston Rsch. Corp. v. Minn. Mining & Mfg. Co., 350 F.2d 134, 142–43 (9th Cir. 1965) (affirming an injunction limited in duration to the time it would have taken the defendant to develop the product independently in a trade secret suit).

[75] *See* 17 U.S.C. § 504(b); Sheldon v. Metro-Goldwyn Pictures Corp., 309 U.S. 390, 402–04 (1940).

[76] *See* 17 U.S.C. § 503(b).

[77] Winter v. Nat. Res. Def. Council, Inc., 555 U.S. 7, 20-21 (2008) (discussing the standard for preliminary relief).

stage discovery protocol, managed under a strict protective order, a practice common in complex technology litigation.[78]

- **Stage One: Documentary Production.** The initial stage would compel the developer to produce a standardized "Provenance Packet." This packet would include the auditable records necessary to trace the model's lineage, such as cryptographically hashed manifests of all training and fine-tuning datasets, any relevant data licenses or attestations, checkpoint lineage graphs, and reproducible build scripts. All materials would be produced subject to a protective order that limits access to the opposing party's outside counsel and designated experts.[79]

- **Stage Two: Forensic Examination by a Neutral Expert.** If the documentary evidence produced in Stage One proves insufficient to resolve the dispute, the court may then appoint a neutral technical expert or special master pursuant to Federal Rule of Civil Procedure 53.[80] This neutral expert would be empowered to conduct sealed, targeted forensic testing of the model itself—for example, to validate a claimed curative rebuild or to probe for the influence of a poisonous predecessor. The expert's findings and conclusions, rather than the raw technical data or model weights, would then be reported to the court and the parties, thereby shielding the underlying trade secrets from direct exposure.

Documentation prepared to satisfy external transparency mandates can serve as valuable corroborative evidence within this protocol. For example, training-data summaries required by the *EU AI Act* or the auditable logs recommended by the *U.S. Copyright Office* can be used in Stage One to verify the completeness and accuracy of a developer's Provenance Packet. While such materials do not themselves determine liability, courts may consider them under protective order to assess the credibility of a developer's lineage claims.

Finally, this framework is backstopped by existing procedural safeguards. Courts may draw adverse inferences or impose sanctions under Federal Rule of Civil Procedure 37(e) where a developer fails to preserve lineage logs, dataset manifests, or checkpoint histories after litigation is reasonably foreseeable. Nothing in this framework expands copyright's territorial reach: the "poisonous tree" determination follows the law governing the act of ingestion, while AI-FOPT allocates evidentiary burdens for models offered in U.S. commerce.

As constructed, these components—trigger, scope, presumption, rebuttal, remedies, and procedure—form a doctrinally coherent and judicially manageable framework. By anchoring the trigger in an adjudicated infringement, the doctrine respects and preserves the central role of fair use analysis at the initial point of ingestion. By defining clear rebuttal paths and providing concrete safe harbors, it allows for continued innovation while ensuring accountability. Its calibrated ladder of remedies ensures that any judicial intervention is proportional to the harm, and its reliance on familiar procedural tools—from statutory burden-shifting analogues to staged discovery under the Federal Rules—equips courts to manage these complex cases without overstepping their existing authority. This framework is thus designed not as a radical departure from copyright law, but as a necessary adaptation to restore its core evidentiary bargain in the age of generative AI. Having detailed the operational mechanics of the AI-FOPT doctrine, the Article now turns to addressing the principal policy objections that such a framework might invite.

---

[78] *See, e.g.*, Order Re Proposed Protective Orders, Oracle Am., Inc. v. Google Inc., No. C 10-03561 WHA (N.D. Cal. Dec. 14, 2010) (setting terms of a protective order governing source-code review and confidentiality, including outside-counsel/experts-only access).

[79] *See* FED. R. CIV. P. 26(c) (authorizing courts to issue protective orders "to protect a party or person from annoyance, embarrassment, oppression, or undue burden or expense").

[80] *See* FED. R. CIV. P. 53 (governing the appointment and powers of a special master).

# IV. POLICY STAKES & OBJECTIONS

The primary justification for adopting an intellectual-property analogue of the FOPT doctrine lies in its potential to restore enforceability and deterrence to copyright law in scenarios of multi-generational AI development. By focusing on the legitimacy of training data lineage, it directly confronts the problem of technological obfuscation, making it harder for infringing practices at the foundational stages of AI development to be shielded by layers of subsequent processing. This approach would deter the initial mass ingestion of unauthorized data for training commercial AI models by making the downstream exploitation of such models legally perilous, thereby encouraging more ethical data sourcing from the outset and upholding the core incentives of copyright articulated in foundational cases.[81]

Nevertheless, an AI-FOPT presumption will invite at least two principal objections. *First*, critics will warn that automatically tainting downstream models will unduly “chill innovation,” deterring legitimate experimentation with machine learning by making AI development prohibitively risky and complex. *Second*, skeptics will argue that a lineage-based rule subverts the flexibility prized in fair-use analysis by cases like *Campbell v. Acuff-Rose Music, Inc.* and *Authors Guild v. Google*, by appearing to privilege process over the final, potentially transformative output.

**Limiting Principles and Safe Harbors**

- **Trigger is narrow:** AI-FOPT attaches only after a court *adjudicates* an upstream training act as infringing (and not excused by fair use). No presumption arises absent that finding.
- **Rebuttable, not strict:** The presumption shifts burdens because provenance evidence sits uniquely with developers; it can be overcome by (i) *clean lineage* or (ii) *purged taint* (curative rebuild or verifiable unlearning).

[81] *See, e.g.,* Sony Corp. of Am. v. Universal City Studios, Inc., 464 U.S. 417, 429 (1984) (discussing balance in copyright); Eldred v. Ashcroft, 537 U.S. 186, 211-15, 227 (2003) (on promoting progress).

- **No expansion of subject matter:** The doctrine does *not* protect “style,” ideas, or systems, and does not impose output strict liability.
- **Remedies are calibrated:** Relief is component-targeted where feasible, consistent with *eBay*. Model destruction is exceptional, not default.
- **Fair use remains primary:** The fair-use analysis at the initial ingestion stage is unchanged and dispositive for the “poisonous tree” determination.
- **Reliance & innocent users:** Courts can tailor relief to reliance interests and culpability. Damages for innocent infringers may be reduced and prospective, component-level remedies can mitigate disruption where equities favor narrower relief.

The concern that an AI-FOPT standard would “chill innovation” misunderstands its function and the nature of sustainable technological progress. Copyright doctrine has long tolerated some burden-shifting and process-focused inquiries to curb systematic infringement, particularly where plaintiffs face profound information asymmetries. Contributory infringement jurisprudence, for example, places an affirmative duty on defendants who materially facilitate third-party copying to take *“reasonable steps”* to prevent unlawful uses once placed on notice.[82] An AI-FOPT standard simply extends that logic up the supply chain: the developer that made the architectural choice to build upon a tainted foundation (an $AI_1$ trained on infringing data) must demonstrate clean provenance for its downstream models or accept the commercial risk. Far from freezing innovation, such a regime would *re-orient it toward lawful inputs*.

While provenance tracking introduces compliance costs, these costs internalize externalities currently imposed on copyright holders and are likely to decrease as standardized tools and “clean-corpus” certifications emerge.[83] The machine-learning community already invests heavily in techniques like *federated learning*, *privacy-preserving synthetic data generation*, and *machine un-learning*, precisely because direct ingestion of personal or copyrighted data is increasingly recognized as ethically problematic and legally risky.[84]

---

[82] Gershwin Publ’g Corp. v. Columbia Artists Mgmt., Inc., 443 F.2d 1159, 1162 (2d Cir. 1971); *see also* MGM Studios, Inc. v. Grokster, Ltd., 545 U.S. 913, 936-37 (2005) (inducement liability where defendant takes affirmative steps to foster infringement).

[83] A related concern is incumbency: that provenance controls and rebuild costs may be easier for large firms to absorb. Two features of the AI-FOPT doctrine mitigate this risk: (1) its rebuttable posture, which allows targeted component-level cures rather than wholesale rebuilds, and (2) the availability of standardized, certifiable “clean-corpus” pathways that reduce fixed compliance costs for smaller developers. Indeed, the current legal *uncertainty*—characterized by massive, bet-the-company litigation—may chill innovation, especially for smaller players, more than a clear, albeit demanding, evidentiary standard.

[84] *See, e.g.,* Peter Kairouz et al., *Advances and Open Problems in Federated Learning*, 14 Found. Trends Mach. Learn. 1, 5–17 (2021) (privacy and data-minimization motivations for FL); M. Gong et al., *A Survey on Differentially Private Machine Learning*, 15 IEEE Comput. Intell. Mag. 49, 52–53 (2020) (surveying DP methods, including DP synthetic-data

Requiring verifiable pedigrees for training corpora would accelerate these technical solutions, not halt them.[85] History suggests that when the law removes the cheapest—but illicit—input, industry rapidly discovers lawful substitutes; the post-*Napster* rise of licensed music streaming services is instructive in this regard.[86] An AI-FOPT principle, by providing clearer (albeit stricter) rules, would reduce legal uncertainty and channel innovation towards ethically sourced data and genuinely transformative uses, fostering a more sustainable and equitable AI ecosystem.

The second objection goes to doctrinal harmony: does FOPT displace fair-use flexibility?

Nothing in the AI-FOPT framework forecloses or undermines *fair use*; it merely *anchors each fair use analysis to a distinct act of ingesting human-authored copyrighted data*. Specifically, for any given model—such as an initial model $AI_1$—fair use remains central to determining whether its training renders it a "poisonous tree." If the model's training is found to involve non-copyrighted data, a legitimate fair use of copyrighted data (as in *Authors Guild v. Google* for its specific transformative purpose of creating a search index[87]), or another applicable legal exception (like the EU's TDM provisions), then the model is not deemed a "poisonous tree." No AI-FOPT presumption of taint arises for the model's outputs or for any subsequent models trained on its outputs. However, if any of these subsequent models *themselves ingest new human-authored copyrighted data*, then each such act of ingestion is subject to its *own* independent fair-use assessment.

Conversely, if a model's training *fails* fair use scrutiny, it is deemed a "poisonous tree." This would occur, for instance, if a developer scraped millions of novels to build a synthetic-text generator aimed at the same readership as the originals.[88] Consequently, the model's outputs become presumptively tainted "fruit," and any subsequent models ($AI_2$, $AI_3$, etc.) trained on those outputs inherit the presumption of taint. Moreover, where its successor models also ingest

---

approaches, for privacy-preserving training); Weiqi Wang et al., *Machine Unlearning: A Comprehensive Survey*, ARXIV:2405.07406v2, at 1–2 (July 25, 2024), https://arxiv.org/abs/2405.07406 [https://perma.cc/CNV9-7CEQ] (overview of unlearning techniques responding to GDPR/"right to be forgotten" and related concerns).

[85] For instance, investments in federated learning have surged in response to privacy laws, yielding robust, compliant AI systems.

[86] *See* Michael A. Carrier, *Copyright and Innovation: The Untold Story*, 2012 WIS. L. REV. 891, 916–19 (2012) (discussing post-Napster licensing dynamics and innovation); *see also* A&M Records, Inc. v. Napster, Inc., 239 F.3d 1004, 1019–23 (9th Cir. 2001) (rejecting a fair-use defense for P2P file-sharing).

[87] Authors Guild v. Google, Inc., 804 F.3d 202, 214–19 (2d Cir. 2015).

[88] A scenario where the first (purpose and character of use, especially commercial and non-transformative in the *Warhol* sense) and fourth (effect on the market for the original) fair use factors, as emphasized in *Campbell*, would *strongly weigh* against fair use for $AI_1$'s training under the first and fourth factors as understood after *Warhol*. *See* Campbell v. Acuff-Rose Music, Inc., 510 U.S. 569, 579-81, 586-94 (1994); Andy Warhol Found. for the Visual Arts, Inc. v. Goldsmith, 143 S. Ct. 1258, 1271–75 (2023).

new, potentially infringing copyrighted data, they may face both this inherited FOPT presumption (stemming from the initial infringement) *and* separate infringement claims (with corresponding fair-use analyses) for each new ingestion.

By propagating taint, AI-FOPT ensures that an initial failure of fair-use scrutiny cannot be "laundered" away by merely layering levels of technological abstraction.[89] The burden then shifts to the developers of downstream models that relied on the outputs of infringing models to rebut the presumption of taint—whether by demonstrating independent creation, implementing curative retraining, or otherwise establishing a clean lineage. However, AI-FOPT does not, in any of these scenarios, impede access to fair use as a defense where it may today be applicable.

Beyond these core copyright considerations, AI-FOPT operates within established constitutional boundaries. The doctrine is content-neutral, regulating the *means of model creation* and *evidentiary burdens*, not protected speech, thereby aligning with First Amendment principles.[90] Its burden-shifting mechanism satisfies due process, as rebuttable presumptions are permissible where there is a "rational connection" between the proven fact (an adjudicated infringement) and the presumed fact (downstream taint), particularly when defendants uniquely control the evidence.[91] It does not evade Article III's standing requirements; plaintiffs must still show concrete injury, but the presumption operates only after a foundational infringement is adjudicated and with respect to models commercially deployed. And unlike the speaker- and content-based restrictions that triggered heightened scrutiny in *Sorrell v. IMS Health Inc.*,[92] AI-FOPT neither disfavors particular speakers nor restricts dissemination of specific content; it conditions downstream commercial exploitation on provenance the developer uniquely controls.

The practical *administrability* of AI-FOPT is demonstrable. Courts already possess a mature toolkit for managing technically complex discovery. In software-copyright and trade-secret litigation, judges routinely compel the production of source code repositories, build logs, and expert reverse-engineering reports; trade-secret cases often require accelerated "red-team" testing of accused products to determine if misappropriated information was used.[93] Similar mechanisms can be employed to verify lineage and the effectiveness of any claimed curative measures. Nor would an AI-FOPT rule ensnare incidental or *de minimis* inspirations. A plaintiff must first make a credible showing of an adjudicated (or demonstrably) infringing $AI_1$ and a model distillation chain leading to the defendant's model. A developer who merely fine-tunes a public-domain language model on their own proprietary corporate documents, for example, would likely never confront the presumption. And even when the presumption attaches, defendants would retain the full panoply of equitable defenses, such as laches, estoppel, and, where genuinely appropriate, evidence of independent creation for the specific model in question. Nothing here alters § 512 safe harbors for hosting providers; AI-FOPT addresses *model developers'* provenance obligations in infringement actions, not intermediary liability.

---

[89] While this approach means a downstream model's transformative purpose cannot retroactively cure an upstream infringement, this trade-off is necessary to close the laundering loophole; to hold otherwise would be to grant a single transformative use at the end of a chain the power to sanitize any number of illicit acts that enabled its creation, rendering copyright's protections illusory.

[90] *See* Eldred v. Ashcroft, 537 U.S. 186, 219–21 (2003); Golan v. Holder, 565 U.S. 302, 327–29 (2012); *cf.* United States v. O'Brien, 391 U.S. 367, 377 (1968) (upholding content-neutral regulation of conduct with expressive elements).

[91] *See* Mobile, Jackson & Kan. City R.R. v. Turnipseed, 219 U.S. 35, 43–44 (1910).

[92] *See* Sorrell v. IMS Health Inc., 564 U.S. 552, 563–71 (2011).

[93] *See, e.g.*, Order Re Proposed Protective Orders, Oracle Am., Inc. v. Google Inc., No. C 10-03561 WHA (N.D. Cal. Dec. 14, 2010); Order Re: Forensic Inspection, Brocade Commc'ns Sys., Inc. v. A10 Networks, Inc., No. 10-cv-03428-LHK (N.D. Cal. Jan. 20, 2012) (ordering a neutral expert to conduct forensic testing of the accused products).

AI-FOPT also offers a degree of *international compatibility*. Because AI training and deployment are often transnational, its trigger is flexible: it imports the legal standards of the jurisdiction where the foundational copying occurred to determine if a model is a "poisonous tree." In the EU, for example, the DSM Directive's text-and-data-mining exceptions (Article 3 for research organizations and cultural-heritage institutions; Article 4 for any purpose subject to rightsholder opt-out) mean that a developer that complies begins with a non-poisonous tree for EU purposes; similar results may follow under fair-dealing regimes such as Canada's.[94] This flexibility does not, however, create a new cause of action for foreign infringement. The doctrine applies only when a U.S. claim can be established based on domestic infringing acts—e.g., reproducing or distributing the model (or its copies) in the United States, publicly performing/displaying outputs in the United States, or importing the model or copies into the United States.[95] This two-step analysis respects legal territoriality while ensuring the evidentiary presumption can be applied consistently to models entering the U.S. market, regardless of their origin.

Finally, critics might argue that FOPT is not a traditional copyright doctrine. While true, this proposal advocates for its *analogical application* to address a novel technological challenge that existing doctrines struggle to manage effectively. Copyright law has long adapted to new technologies. The AI Ouroboros, with its unique systemic risk of rendering copyright unenforceable through technological obfuscation, warrants a similarly pragmatic response.

# V. CONCLUSION: RESTORING THE EVIDENCE MACHINE

Copyright's traditional *access plus similarity* evidence machine has long served human creative endeavors effectively by providing a clear framework to adjudicate infringement based on tangible comparisons. However, as this Article demonstrates, that machine falters when confronted with AI trained on recursive synthetic pipelines—an AI Ouroboros. When successive

---

[94] Directive (EU) 2019/790, arts. 3–4, 2019 O.J. (L 130); CCH Canadian Ltd. v. Law Soc'y of Upper Canada, 2004 SCC 13 (Can.).

[95] *See generally* Subafilms, Ltd. v. MGM-Pathe Commc'ns Co., 24 F.3d 1088, 1095–98 (9th Cir. 1994) (en banc) (presumption against extraterritorial application of the Copyright Act); Morrison v. Nat'l Australia Bank Ltd., 561 U.S. 247, 255–61 (2010) (articulating the modern two-step test); 17 U.S.C. § 602(a)(1) (importation).

AI generations ingest the outputs of their predecessors, they create a technological hall of mirrors where infringement can become both systemic and forensically inscrutable. Original copyrighted works can get buried deep within opaque training processes, and similarity to any single source can become stochastic and diffuse. The evidentiary thread connecting final output to original input effectively snaps, leaving rights holders with a near-insurmountable burden of proof. Left unchecked, this dynamic risks entrenching a perverse equilibrium: systems that profit most from copyrighted expression may also bear the least accountability for its exploitation.

Precedent in copyright and adjacent fields already points toward an antidote. In domains where merely comparing end products fails to capture the essence of the wrong—such as with bootleg master recordings or illicitly created molds for artistic reproduction—courts have cut off liability at its source or focused on the illicit act of taking itself. By tethering liability to training data lineage, AI-FOPT adopts analogous logic. Just as *Wong Sun* barred prosecutors from sanitizing tainted evidence through successive layers of derivations, it seeks to prevent developers from sanitizing copyrighted expression through successive (statistical) abstractions.

It does so while preserving copyright's traditional boundaries: it neither expands the scope of protectable subject matter nor extends the duration of exclusivity. Rather, it enforces existing rights against a novel evasion tactic—one that weaponizes scale, stochasticity, and technical opacity to obscure systemic appropriation, effectively "laundering" copyright.

AI-FOPT is doctrinally sound, judicially manageable, and commercially viable. For courts, it provides an administrable standard grounded in familiar equitable principles. Courts can (1) condition preliminary relief on production of a standardized provenance packet under protective order, (2) appoint a neutral technical master for sealed validations under Fed. R. Civ. P. 53, and (3) tailor component-level injunctions or running royalties consistent with *eBay* and § 504. This lineage-based inquiry parallels chain-of-title analyses in real property and provenance tracing required for cultural artifacts—parallel conceptual anchors, designed to help overcome hurdles relating to abstraction and obscurity. For policymakers, it carves a path between laissez-faire permissiveness and draconian bans, balancing deterrence against the risks of overbreadth. For industry, it incentivizes ethical sourcing through liability rules that scale with harm, potentially catalyzing new markets for licensed datasets. The result could be a self-reinforcing cycle: as "clean-label" datasets and AI models become market differentiators, rights holders may gain leverage to negotiate collective licenses, and developers may gain the stability of clearer legal ground, reducing reliance on adversarial litigation.

The choice need not be between stifling AI and surrendering copyright. It can be between permitting infringement to metastasize behind algorithmic opacity or ensuring the evidentiary accountability necessary for human creativity and machine learning to advance symbiotically. The AI Ouroboros need not devour copyright—the law fundamental to its creation—or the incredible innovation that it can sustain.